# A Self-Updating K-Contingency List for Smart Grid System

Sohini Roy, Arunabha Sen
*School of Computing, Informatics and Decision System Engineering*
*Arizona State University*
Tempe-85281, Arizona, USA
Email: {sohini.roy, asen}@asu.edu

*Abstract*—A reliable decision making by the operator in a smart grid is contingent upon correct analysis of intra-and-interdependencies between its entities and also on accurate identification of the most critical entities at a given point of time. A measurement based self-updating contingency list can provide real-time information to the operator about current system condition which can help the operator to take the required action. In this paper, the underlying intra-and-interdependencies between entities for a given power-communication network is captured using a dependency model called Modified Implicative Interdependency Model (MIIM) [1]. Given an integer K, the event-driven self-updating contingency list problem gives the list of K-most critical entities, failure of which maximizes the network damage at the current time. Owing to the problem being NP complete, a fast heuristic method to generate a real-time contingency list using system measurements is provided here. The validation of the work is done by comparing the contingency list obtained for different K values using the MIIM model on a smart grid of IEEE 14-Bus system with that obtained by simulating the smart grid using a co-simulation system formed by MATPOWER and Java Network Simulator (JNS). The results also indicate that the network damage predicted by both the ILP based solution [2] and the proposed heuristic solution using MIIM are more realistic compared to that obtained using another dependency model called Implicative Interdependency Model (IIM) [3].

## I. INTRODUCTION

The smart grid system can be viewed as a two-layered network where one layer is composed of the power entities and the other layer is formed with communication entities. Yet, both the layers are coupled together by complex interdependencies in order to function as a single system. Moreover, the entities of each layer of the network also exhibits intra-dependencies between them. Such intra as well as interdependencies are critical in a smart grid system as a failure of one or more entities in each layer can lead to a cascading failure of multiple entities in the whole system. Therefore, it is necessary to understand the dependencies between different entities of the smart grid and identify the most critical entities, damage of which can result in the failure of maximum number of entities in the smart grid system. Such entities can be identified as the most vulnerable entities upon which the operability of the smart grid system is contingent and a list of such entities in the system is termed as the contingency list [4].

Appropriate protection of the entities in the contingency list can save the smart grid from a catastrophe. However, in order to identify the contingency list, (i) the design of each layer of network should be clear to the researchers, and (ii) there should be an appropriate model that can represent the complex intra-and-interdependencies between the entities in the smart grid. In [2], a simplistic and easy Boolean Logic based dependency model, termed as the Implicative Interdependency Model (IIM) is proposed which very accurately captures the complex interdependencies between the entities in a critical infrastructure. Yet, IIM also has drawbacks like (i) ignoring the effects of failure of entities in the Information and Communication Technology (ICT) layer, (ii) lack of ICT network design information and (iii) binary nature; that means an entity can either fail or operate only and it does not take into account any reduced level of operation for the entities in the system. In [8], a rough idea about the ICT network for smart grid is provided but it also lacks the detailed description of the ICT design. A Modified Implicative Interdependency Model (MIIM) proposed in [1] takes into account all such drawbacks and provides a clear picture of the smart grid network design by taking inputs from a utility in the U.S. Southwest. It also considers different operational levels of the entities and model the complex dependencies using multi-valued Boolean Logic based equations called Interdependency Relations (IDRs). The contingency list can be obtained for a smart grid system, just by solving these IDRs.

Now, according to [1], the IDRs are updated if any failure or change of operational level occurs to the entities. Therefore, the contingency list for a smart grid in a steady state will also differ from the list when some failure takes place in the system and the IDRs are changed. Thus, an event-driven self-updating contingency list is required to offer resilience to the grid. Again, since the contingency list keeps on changing with any event of failure in the system, it is difficult to harden all the entities in updated the contingency list every time and the operator might have some budget constraints for hardening the entities in the list. Such budget constraint can either be monetary or availability of resources. If the current budget of hardening is K-entities for the operator, then it is essential to identify the K-most vulnerable entities in the network at the current time. This K can be any integer value, less than the total number of entities in the smart grid. In this paper, a novel method to generate a self-updating K-contingency list at for a smart grid is proposed. It is already proved in [2] that the problem of identifying the K-contingency list is NP-complete. Therefore, an Integer Linear Programming (ILP) based solution for the problem is provided in this paper using the MIIM IDRs. Finally, a validation of the results obtained from the proposed method is done by co-simulating the two layers of the smart grid network of IEEE 14-Bus system using MATPOWER and Java Network Simulator (JNS). A comparative study of the K-contingency list obtained using the MIIM IDRs is done with that obtained using IIM for a smart grid of IEEE 14-Bus system also.

The rest of the paper is structured as follows. Section II gives an overview of the IIM and MIIM models. Problem definition of the self-updating K-contingency list is given in Section III. Section IV gives a fast heuristic solution for the problem. A comparative analysis between the K-contingency lists obtained using MIIM, IIM ILP and heuristic solutions with smart grid network layers co-simulation results is provided in Section V. Section VI concludes the paper.

## II. OVERVIEW OF IIM AND MIIM

In both IIM [2] and MIIM [1], the smart grid system can be viewed as a multilayer network, represented as a set $J(E, F(E))$, where $E$ represents set of all entities in both the layers of the smart grid and $F(E)$ represents the set of IDRs. The entities in power layer (layer 1) are considered as P type entities where $P = \{P_1, P_2, \ldots P_n\}$ and entities in ICT layer (layer 2) are named as C type entities where $C = \{C_1, C_2, \ldots C_m\}$. The set $F(E)$ is used in both the models to capture the dependencies among interacting entities in the network. Yet, only structural dependencies are considered to generate the IDRs in IIM and both structural as well as operational aspects of the entities are taken into account while formulating IDRs for MIIM. IIM has a binary nature and the entities in that model can either be operational with a state value of 0 or be non-operational with a state value of 1. The most common feature of reduced operability in critical infrastructures is ignored in IIM. The entities in MIIM can take a value of 0, 1 and 2 indicating no-operation, reduced operation and full operation respectively.

Let $C_i$, an entity of layer 2, be operational if (i) $C_j$ which is another entity of layer 2 and $P_a$ which is an entity of layer 1, are operational, or (ii) $C_k$ which is an entity of layer 2 and $P_b$ which is an entity of layer 1 are operational, and (iii) $C_l$ which is an entity in layer 2 is operational. Then the corresponding IIM IDR for $C_i$ would be: $C_i \leftarrow \left((C_j . P_a) + (C_k . P_b)\right) . C_l$. In this IDR, '.' denotes logical AND operation and '+' denotes logical OR operation. Similarly, the IDR for a P type entity can be expressed.

In MIIM, three Boolean operators are used while formulating the IDRs. The first operator is min-AND, denoted by '○', which selects the lowest of its input values. The second operator is max-OR, denoted by '●', which selects the highest of its input values. The third operator is new_XOR, which is denoted by '◉'. If all the inputs of new_XOR are same, then the output is also same as the inputs. In all other cases the output is 1. This new_XOR operator actually denotes the level of operation of an entity. The truth table for all the 3 new operators are given in Table I.

TABLE I. TRUTH TABLE FOR MIIM OPERATORS

| Input 1 | Input 2 | min-AND | max-OR | new_XOR |
|---|---|---|---|---|
| 2 | 2 | 2 | 2 | 2 |
| 2 | 1 | 1 | 2 | 1 |
| 2 | 0 | 0 | 2 | 1 |
| 1 | 2 | 1 | 2 | 1 |
| 1 | 1 | 1 | 1 | 1 |
| 1 | 0 | 0 | 1 | 1 |
| 0 | 2 | 0 | 2 | 1 |
| 0 | 1 | 0 | 1 | 1 |
| 0 | 0 | 0 | 0 | 0 |

TABLE II. EVALUATION OF IIM AND MIIM IDRs

|  | IIM | MIIM |
|---|---|---|
| STEP 1 | $C_l \rightarrow 0$ | $C_l \rightarrow 0$ |
| STEP 2 | $C_i \leftarrow (((2.2) + (2.2)).0)$ | $C_i \leftarrow (((2 \circ 2) \bullet (2 \circ 2)) \circledcirc 0)$ |
| STEP 3 | $C_i \leftarrow ((2 + 2).0)$ | $C_i \leftarrow ((2 \bullet 2) \circledcirc 0)$ |
| STEP 4 | $C_i \leftarrow (2.0)$ | $C_i \leftarrow (2 \circledcirc 0)$ |
| STEP 5 | $C_i \leftarrow 0$ | $C_i \leftarrow 1$ |

In order to illustrate MIIM, let us assume that if an entity in condition (i) or (ii) fails, $C_i$ will still work full operability, but if (iii) is not satisfied then $C_i$ will operate at a reduced level; this relation can be expressed using MIIM IDRs as: $C_i \leftarrow \left((C_j \circ P_a) \bullet (C_k \circ P_b)\right) \circledcirc C_l$. To differentiate between the two models in terms of smart grid system application, the failure of entity $C_l$ for the above IIM and MIIM IDRs are considered and the outcomes are observed in Table II.

It is observed in Table II, that for same kind of dependencies, failure of the entity $C_l$ results in the failure of entity $C_i$ in case of IIM but it only reduces the operation level in case of MIIM.

## III. SELF-UPDATING K-CONTINGENCY LIST PROBLEM

The operator of a smart grid system relies on the sensor-based data like PMU-data and RTU-data to know about the operational state of each and every entity in the power grid. Therefore, it is equally important for the operators to know about the operational states of the communication entities carrying data from the sensors placed in the substations to the control centers. If the operational level of an entity in the system reduces then immediate actions can be taken by the operator. Hence, at a real time, the entities which are more vulnerable to failure should be identified and proper protection or backup to those entities should be provided. This calls the need for an automated system generating the K-Contingency List for the current smart grid system, so that the maximum damage in the power-communication network can be avoided. When one or more entities fail in the smart grid system, many other entities also fail as a result and this is called cascading failures, and this often might lead to a catastrophe if not arrested in time. This cascade stops when the system reaches a steady state once again. Each time a failure takes place in the smart grid, the set $J(E, F(E))$ is updated. All entities that get a state value 0 are removed from the set $E$. As a result, all the IDRs in set $F(E)$ are also updated, since all the dependencies with those failed entities are removed. Now, in between two steady states of the system, there are a number of unstable states of the smart grid when the cascade propagates. Propagation of this cascade may not take place instantly and therefore measures can be taken to arrest the cascade by identifying the K-Contingency List at that time. Given an integer K, and a smart grid system represented as set $J(E, F(E))$, this problem returns the set of K-most critical entities in the joint network, failure of which can lead to the maximum total number of failed entities in the system at the end of the cascade propagation. It is to be noted that a cascade can only propagate in one direction since an already failed entity cannot be affected again by the cascading failure. Therefore, upper bound of the cascade is $|EG| - 1$; where EG is the total number of edges in the network. A formal definition of the problem using the MIIM [1] model is as follows:

### A. Inputs to the Problem

- (a) A joint network $J(E, F(E))$; where $E = P \cup C \cup CP$

    - $P = B \cup T \cup Batt$ (Buses, Transmission Lines/Transformers, Batteries)
    - $C = SE \cup SRE \cup DRE$ (Substation Entities, SONET-Ring Entities, DWDM-Ring Entities)
    - $CP = L \cup R \cup U$ (Power supply lines, RTUs and PMUs)

- (b) Two positive integers K and S.

### B. Decision version of the Problem

Does there exist a set of K entities in E whose failure at time t would result in a failure of at least S entities in total at the next state of the cascading process?

### C. Optimization version of the Problem

Compute the set of K entities in the joint network $J(E, F(E))$ whose failure at time t would maximize the number of entities failed or in other words minimize the overall system state values in the next state of cascade propagation.

The problem of finding K-Contingency List is NP complete, which is already proved in [3]. Therefore, an ILP based solution for the problem is given in [2] and a faster heuristic solution is given in Section IV of this paper. Also, validation of the results is done by comparing the ILP based and heuristic solution results with the simulation results.

## IV. HEURISTIC SOLUTION USING MIIM IDRs

The heuristic solution to the self-updating K-Contingency list is completely based on the observations made during the ILP based solutions and simulations.

In order to solve the problem heuristically, first the smart grid system should be considered as a graph $G = (V_P, V_C, E_{PC}, E_{PP}, E_{CC})$ consisting of two different types of vertices $V_P$ and $V_C$ and three different types of edges $E_{PC}, E_{PP}$ and $E_{CC}$. In this abstraction, $V_P$ indicate the power network buses and $V_C$ indicate the communication entities except the channels. All the power or communication channels that connect power and communication entities eg: power supply lines to the communication entities are denoted by $E_{PC}$, Transmission lines and transformers are denoted by $E_{PP}$ and all communication channels are denoted by $E_{CC}$. We are assuming that any edge cannot be most critical as all power networks are (n-1) fault tolerant and all communication networks can adjust routing technique based on failed channels.

*1) Initially all the vertices in the graph are considered to be white in color.*

*2) Input:* $G = (V_P, V_C, E_{PC}, E_{PP}, E_{CC})$, K, set of MIIM IDRs and a state table having the state values of each entity.

*3) Step 1:* The $V_P$ vertices corresponding to generator buses in the actual grid are identified and colored yellow.

*4) Step 2:* The $V_P$ vertices corresponding to buses with a PMU in the actual grid are identified and colored blue. Any $V_P$ satisfying both the criteria of Step 1 and 2 will be green in color.

*5) Step 3:* Step 3 will solve the problem for K=1.
- Consider a subgraph $G_1 = (V_P, E_{PP})$ ; since a failure of any communication entity cannot bring maximum damage to the smart grid.
- If the graph has pendant vertices:
    - Identify the pendant $V_P$ vertices.
    - Identify the $V_P$ vertices connected to those pendant vertices and color them Pink.
- Else if the graph does not have pendant vertices:
    - Identify the $V_P$ vertices having minimum connections.
    - Color those nodes pink.
- Check the total damage caused by failure of each such pink node by solving MIIM IDRs for those entities only.
- Select the nodes resulting in maximum damage and color them red.
- A list of all such red nodes comprise the K=1 contingency list.
- Change all pink nodes to their previous color.

*6) Step 4:* Step 4 will solve the problem for K=2.
- Take two empty lists List1 and List2.

- Consider a subgraph $G_1 = (V_P, E_{PP})$; since a failure of just two communication entities cannot bring maximum damage to the smart grid.
- Combine each of the red nodes to each of blue, green and yellow nodes to form all pairs of {Red, Green}, {Red,Yellow} and {Red, Blue}.
- Check the total damage caused by failure of each such pair by solving MIIM IDRs for those entities in each pair only.
- Find the {Red, G/Y/B} pair(s) failure of which causes the maximum damage.
    - Add the pair(s) in List1
- Find all $V_P$ vertices having two $E_{PP}$ edges only.
- Identify the $V_P$ vertices connected to such $V_P$ vertices having two $E_{PP}$ edges only.
- Color all such $V_P$ vertices grey.
- For all such pair of grey $V_P$ vertices:
    - Check the total damage caused by the pair
    - Find the pair(s) causing maximum damage.
    - Add the pair(s) in List2
- Compare the total damage caused by List1 pairs and List2 pairs
- Change all grey nodes back to their previously assigned color.
- All the pairs causing maximum damage, comprise of the K=2 contingency list.

*7) Step 5: If K>2, this step is executed*
- Round =0, TList1=Empty, TList2=Empty (Round is a counter and TList1 and Tlist2 are two temporary lists)
- KCon_List =Empty (KCon_List is the K-Contingency List)
- Graph G2←G1
- While (TList2 is Empty)
    - Find the list of K=2 most vulnerable entities in a list named List_Round (Using step 4 and the input graph G2)
    - Remove all the entities in List_Round from the graph G2 and all the connections associated with them.
    - Add the pairs in TList1
    - If the number of pairs in TList1>=K/2
        - Find all combinations of the pairs in TList1 resulting in a K set.
        - Check the K set causing maximum damage using MIIM IDRs.
        - TList2 ← all such K sets.
- If K is Even
    - KCon_List←TList2
- Else
    - Consider graph (G1-{Entities in TList2})
    - Convert all the previous red nodes to their last assigned colors.
    - Repeat step 3
    - Combine TList2 with each current red node obtained.
    - Check the damage caused by solving MIIM IDRs
    - Find all combinations of TList2 and Red node causing maximum damage
    - KCon_List ← Each such combinations

*8) Step 6:* Check if any new failure takes place in the system.
- If yes
    - Update the state values in state table.
    - Remove IDRs of those entities.
    - Remove the entities from the input graph.
    - Repeat step 3 to 6.
- If No
    - Check if there are $V_C$ vertices having all edges $E_{PC}$ connecting them to the $V_P$ entities in the failed list.
    - If yes:
        - Color such $V_C$ vertices red
        - Add such $V_C$ vertices in the K=1 contingency list.
    - Check if there are $V_C$ vertices having all edges $E_{CC}$ connecting them to the $V_C$ entities in the failed list.
    - If yes:
        - Color such $V_C$ vertices red
        - Add such $V_C$ vertices in the K=1 contingency list.
    - The $V_C$ vertices having all edges $E_{PC}$ connecting them to the $V_P$ entities in the contingency list, are also colored red and added to the K=1 contingency list.
    - The $V_C$ vertices having all edges $E_{CC}$ connecting them to the $V_C$ entities in the contingency list, are also colored red and added to the K=1 contingency list.

The main goal of the heuristic solution of the self-updating K-Contingency list is to reduce the search space in order to reduce the computation time of the problem.

## V. COMPARATIVE ANALYSIS BETWEEN IIM, MIIM AND SIMULATED RESULTS

The smart grid of IEEE 14-Bus system is considered for analyzing the simulation results. The same co-simulation platform using MATPOWER and Java Network Simulator can be used to simulate larger networks also but finding the K-Contingency list for larger networks is difficult as the problem is NP complete. The nomenclature followed for the entities in the smart grid is same as in [1].

### A. Number of entities in the K=1 contingency list Vs. Time (for initial failure of $P_{12}$)

After bus $P_{12}$ located in substation 6 of the smart grid of IEEE 14-Bus fails initially, the contingency list of the system for the next few seconds is analyzed using the MIIM based ILP and heuristic solutions, IIM based ILP solution and the co-simulation method in fig.1.

It is observed that, the simulation results also give the same contingency list as MIIM. It is assumed that no new failures take place even after 5 ms of the failure of bus $P_{12}$. Based on the value of K (1 in this case), the most vulnerable entities in the contingency list are selected.

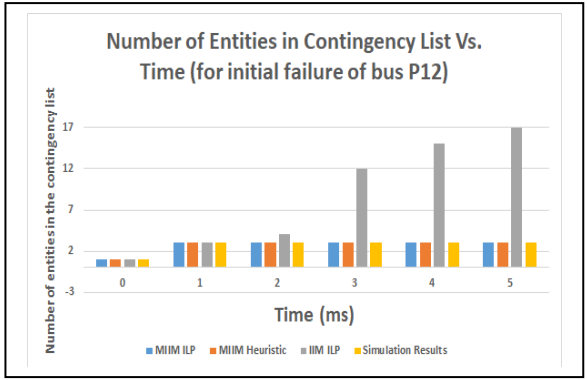

Fig. 1. Number of entities in the contingency list Vs. Time (for initial failure of $P_{12}$)

The Table III shows the entities in the K=1 contingency list for MIIM and IIM after $P_{12}$ fails:

TABLE III. SELF-UPDATING CONTINGENCY LIST

| T (ms) | MIIM Contingency List | IIM Contingency List |
|---|---|---|
| 0 | $P_{12}$ fails | $P_{12}$ fails |
| 1 | $\{P_7\},\{C_{1,2,6,6}\},\{C_{1,1,6,6}\}$ | $\{P_7\},\{C_{1,2,6,6}\},\{C_{1,1,6,6}\}$ |
| 2 | $\{P_7\},\{C_{1,2,6,6}\},\{C_{1,1,6,6}\}$ | $\{P_7\},\{C_{1,2,6,6}\},\{C_{1,1,6,6}\},\{C_{2,1,1,0}\}$ |
| 3 | $\{P_7\},\{C_{1,2,6,6}\},\{C_{1,1,6,6}\}$ | $\{P_7\},\{C_{1,2,6,6}\}$ , $\{C_{1,1,6,6}\},\{C_{2,1,1,0}\}$ , $\{C_{1,2,7,7}\}$ , $\{C_{1,2,8,8}\},\{C_{1,2,9,9}\},\{C_{1,2,11,11}\},\{C_{1,1,7,7}\},\{C_{1,1,8,8}\},\{C_{1,1,9,9}\},\{C_{1,1,11,11}\}$ |
| 4 | $\{P_7\},\{C_{1,2,6,6}\},\{C_{1,1,6,6}\}$ | $\{P_7\},\{C_{1,2,6,6}\}$ , $\{C_{1,1,6,6}\},\{C_{2,1,1,0}\}$ , $\{C_{1,2,7,7}\}$ , $\{C_{1,2,8,8}\}$ , $\{C_{1,2,9,9}\}$ , $\{C_{1,2,11,11}\},\{C_{1,1,7,7}\},\{C_{1,1,8,8}\},\{C_{1,1,9,9}\},\{C_{1,1,11,11}\},\{C_{3,1,1,0}\},\{C_{3,1,4,0}\},\{C_{3,1,5,0}\}$ |
| 5 | $\{P_7\},\{C_{1,2,6,6}\},\{C_{1,1,6,6}\}$ | $\{P_7\},\{C_{1,2,6,6}\}$ , $\{C_{1,1,6,6}\},\{C_{2,1,1,0}\}$ , $\{C_{1,2,7,7}\}$ , $\{C_{1,2,8,8}\}$ , $\{C_{1,2,9,9}\}$ , $\{C_{1,2,11,11}\},\{C_{1,1,7,7}\},\{C_{1,1,8,8}\},\{C_{1,1,9,9}\},\{C_{1,1,11,11}\},\{C_{3,1,1,0}\},\{C_{3,1,4,0}\},\{C_{3,1,5,0}\},\{C_{1,2,10,10}\},\{C_{1,1,10,10}\}$ |

### B. Number of entities in the K=1 contingency list Vs. Time (for initial failure of $P_1$ and $P_{12}$)

Fig.2 shows the contingency list for MIIM ILP, MIIM Heuristic, IIM ILP and Simulated result after P1 and P12 fails initially and no new failures take place even after 8 milliseconds. It is observed that the simulated result of contingency list is same as that obtained using MIIM ILP and MIIM Heuristic. The results obtained using IIM ILP differ a lot from the simulated contingency list. This validates the MIIM model and the heuristic solution proposed in this paper.

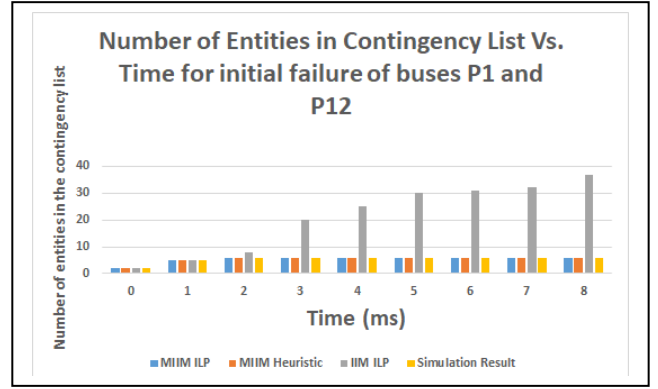

Fig. 2. Number of entities in the contingency list Vs. Time (for initial failure of $P_1$ and $P_{12}$)

### C. Maximum Entities damaged vs. K value

In fig.3., the maximum damage to the network after the initial failure of K-most vulnerable entities are predicted by the ILP based solution to the problem using MIIM IDRs and IIM IDRs. Result obtained by solving the problem heuristically using MIIM IDRs is also shown in the figure. The predicted damages are compared with the simulated results for a smart grid system of IEEE-14Bus.

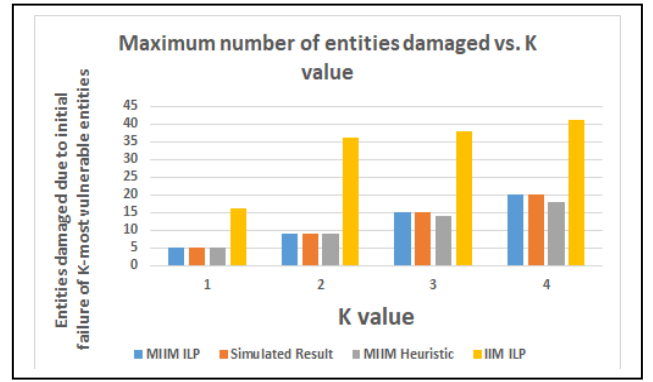

Fig. 3. Maximum number of entities damaged due to the initial failure of K-most vulnerable entities vs. K value

The time taken in generating the ILP based result is compared with that in generating the heuristic result and simulated result is shown in fig.4 and fig.5. In fig.4 and fig.5, the time taken in generating IIM results are not shown as it is already proved from fig.3 that the results are very unrealistic.

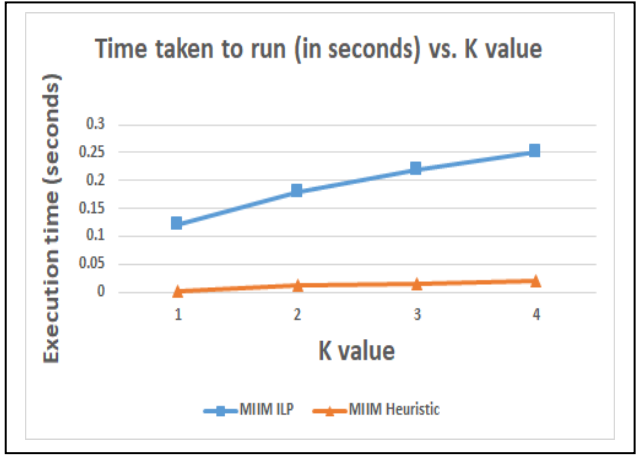

Fig. 4. Time taken to generate the MIIM ILP and Heuristic Solution

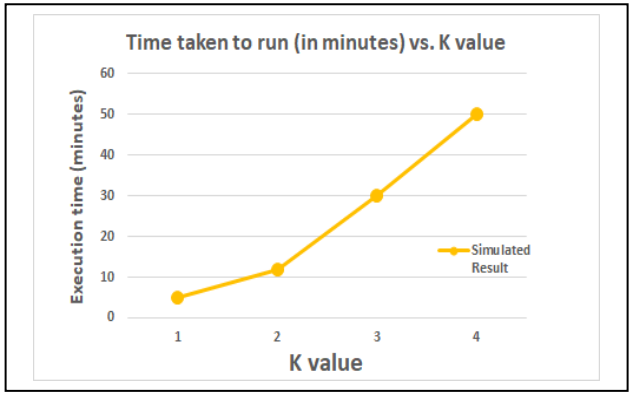

Fig. 5. Time taken to generate the Simulation Result

## VI. CONCLUSION

Using a dual platform based co-simulation of the power and communication network of the smart grid system to verify the interdependency model MIIM is a novel approach proposed in this paper. The MIIM ILP based solution is verified here using this co-simulation. Since the problem is NP-complete, only a small smart grid system of IEEE 14-Bus is considered for performing the simulation. Yet, since the ILP based solution is verified using the co-simulation, the K-most vulnerable entities for any huge smart grid system can be easily obtained in a short time by using the ILP based solution with the MIIM IDRs. Also, the heuristic solution provides results that are very close to the ILP based solution in a much faster way. Therefore, this faster heuristic approach can be used in real smart grids with time constraint, in order to obtain a self-updating K-Contingency list just by updating IDRs after any failure takes place in the smart grid. Applying the MIIM based ILP solution and the heuristic solution for larger smart grid systems is a scope of future work.